\documentstyle[11pt]{article}

  \evensidemargin=\oddsidemargin
\def\title#1{\uppercase{#1}}
\def\abstract#1{\centerline{ABSTRACT} \begin{quotation} #1 \end{quotation}}
\def\beginpaper{\relax}
\def\endpaper{\relax}

\def\beginapjbib{\begingroup \section*{References}
         \parskip=.5ex plus 1.0pt
	 \def\bibitem{\par \noindent \hangindent\parindent
		\hangafter=1}}
\def\endapjbib{\par \endgroup}

\def\be{\begin{equation}}
\def\ee{\end{equation}}
\def\<{\left<}
\def\>{\right>}
\def\sci#1{\ifmmode \times 10^{#1} \else $\times 10^{#1}$ \fi}
\def\Pksq{\ifmmode \bigl< \left| \delta_k \right|^2 \bigr> \else
  $\bigl< \left| \delta_k \right|^2 \bigr>$\fi}

\def\vec#1{\hbox{\boldmath $#1$}}
\def\hii{H~{\sc ii}}
\def\kmax{\ifmmode k_{\rm max} \else $k_{\rm max}$ \fi}
\def\etamax{\ifmmode \eta_{\rm max} \else $\eta_{\rm max}$\fi}
\def\lambdamin{\ifmmode \lambda_{\rm min} \else $\lambda_{\rm min}$\fi}
\def\ddash{\,\hbox{--}\,}
\def\erf{\mathop{\rm erf}}

\def\Mpc{\;{\rm Mpc}}
\def\kpc{\;{\rm kpc}}
\def\etal{{\em et al.}}
\def\la{\mathrel{\mathpalette\fun <}}
\def\ga{\mathrel{\mathpalette\fun >}}
\def\fun#1#2{\lower3.6pt\vbox{\baselineskip0pt\lineskip.9pt
  \ialign{$\mathsurround=0pt#1\hfil##\hfil$\crcr#2\crcr\sim\crcr}}}

\def\element#1#2{\hbox{${}^#2{\rm #1}$}}
\def\D{{\rm D}}
\def\H{{\rm H}}
\def\He#1{\element{He}{#1}}
\def\Li#1{\element{Li}{#1}}

\def\beginfigures{\newpage \section*{Figures} \begin{enumerate}}
\def\endfigures{\end{enumerate}}
\def\figure#1{\item \label{#1}}

\begin{document}

\begin{center}
\bigskip
\rightline{FERMILAB-Pub-94/???-A}
\rightline{UMN-TH-1307/94}
\rightline{astro-ph/9410007}
\rightline{submitted to {\it The Astrophysical Journal}}

\vspace{.35in}

\title{Large Scale Baryon Isocurvature Inhomogeneities}
\bigskip

\vspace{.2in}
Craig J. Copi,$^1$ Keith A. Olive,$^2$ and David N. Schramm$^{1,3}$

\vspace{.2in}
{\it ${}^1$The University of Chicago, Chicago, IL~~60637-1433}\\

\vspace{0.1in}
{\it ${}^2$School of Physics \& Astronomy \\ University of Minnesota,
Minneapolis, MN~~55455}

\vspace{0.1in}
{\it $^3$NASA/Fermilab Astrophysics Center\\
Fermi National Accelerator Laboratory, Batavia, IL~~60510-0500}\\
\end{center}

\medskip

\abstract{Big bang nucleosynthesis constraints on baryon isocurvature
perturbations are determined.  A simple model ignoring the effects of the
scale of the perturbations is first reviewed.  This model is then extended to
test the claim that large amplitude perturbations will collapse, forming
compact objects and preventing their baryons from contributing to the observed
baryon density.  It is found that baryon isocurvature perturbations are
constrained to provide only a slight increase in the density of baryons in the
universe over the standard homogeneous model.  In particular it is found that
models which rely on power laws and the random phase approximation for the
power spectrum are incompatible with big bang nucleosynthesis unless an {\em
ad hoc}, small scale cutoff is included.}

\newpage

\beginpaper

\section{Introduction}

	Big bang nucleosynthesis (BBN) has produced well studied predictions
of the light element abundances (Yang \etal~1984; Walker
\etal~1991, hereafter WSSOK; Smith, Kawano, \& Malaney~1993; Kernan \&
Krauss~1994; Copi, Schramm, \& Turner~1994).  These predictions restrict the
total baryonic contribution to the critical density of the universe,
$\Omega_B$, to be $\Omega_B \la 0.1$.  There have been many attempts to get
around this bound and to extend it to the theoretically preferred value
$\Omega_0 = 1$.  A notable class of such attempts is inhomogeneous BBN (for a
review see Malaney \& Mathews~1993).  These studies of small scale
inhomogeneities in the neutron to proton ratio including hydrodynamic effects,
diffusion, extended networks, and multizone calculations turn out, however, to
provide no appreciable increase on the bound to $\Omega_B$ set by standard BBN
(Kurki-Suonio et al., 1990; Mathews et al., 1990; Terasawa \& Sato, 1991;
Thomas et al. 1994; Jedamzik, Fuller, \& Mathews,~1994)\@.

	On another front, structure formation theories are being constrained
by a rapidly growing body of observational data.  From this the primeval
isocurvature baryon~(PIB) model has fared relatively well.  The PIB model
relies solely on baryons to make up the matter in the universe and
isocurvature perturbations to generate the structure (Peebles~1987a,b; Cen,
Ostriker, \& Peebles~1993).  One short coming of this theory is that it
requires $\Omega_0 = \Omega_B = 0.1 \ddash 0.2$, above the upper bound on
$\Omega_B$ from BBN\@.  Independently, it has been suggested that non-linear
isocurvature fluctuations may allow a larger contribution by baryons than
allowed for in the standard, homogeneous case (Hogan~1978; Hogan~1993).

	In this work we have looked at the effect of large scale isocurvature
perturbations on nucleosynthesis.  Our treatment follows that of Epstein
\& Petrosian (1975, hereafter EP) and Yang et al. (1984). In those efforts,
 it was assumed that the volume distribution for the nucleon abundance
could be described by a gamma distribution in the baryon to
photon ratio, $\eta$. The abundances of the light elements can then be used
to constrain the parameters of the nucleon abundance distribution.
Here we will update that analysis utilizing the most recent constraints
available from the light elements including \Li7.  We will also
consider additional forms for the nucleon abundance distribution
to include the log normal (Barrows \& Morgan~1983) and
the gaussian (Sale \& Mathews~1986) distributions.  Furthermore we have
extended the analysis to include a distribution of power on different scales
to allow for dense regions to form compact objects and hence not contribute
their light elements to the observed abundances.

	For our model we assume the perturbations can be described by a power
spectrum with random phases.  We have no knowledge of the spatial
distribution, $\eta(\vec x)$, and instead specify the choice of the density
probability distribution, $f(\eta)$.  Recently Gnedin, Ostriker, \&
Rees~(1994) have also considered the effects of baryon perturbations on
nucleosynthesis.  They chose a log-normal density distribution with two
scales.  We choose three different density distributions (including the
log-normal distribution) defined on a single scale. We have not specifically
chosen parameters to match those of the PIB model.  Our results, where
comparable, agree with theirs.  Gnedin, Ostriker \& Rees (1994) also
approached the problem from the opposite direction; they assumed a form for
$\eta (\vec x)$ and derive a density distribution $f(\eta)$.  They have found
some models that can circumvent our bounds at the expense of assuming
correlated phases.

	An outline of the paper is as follows: in \S 2 we discuss the
observational bounds on the light element abundances used in this paper, in
particular, how they differ from the values found in WSSOK\@.  In \S 3 we
describe our model for the inhomogeneities.  In \S 4 we present the results
of our calculations.

\section{Observational Limits}

	Observational measurements of the light element abundances play the
crucial role of constraining the standard big bang nucleosynthesis
model as well as models of nucleosynthesis which include inhomogeneities.
 The process of
extracting abundances from the measurements, in particular primordial
abundances, is a difficult task.  An analysis of this process in the context
of limits on BBN is given in WSSOK\@.  The 95\% confidence limit ($2\sigma$)
primordial abundances quoted in WSSOK are
\begin{eqnarray}
	Y_P & = & 0.23 \pm 0.01, \nonumber \\
	\D/\H & \ge & 1.8 \sci{-5}, \nonumber \\
	( \D + \He3 )/ \H & \le & 1.0\sci{-4}, \nonumber \\
	\Li7/\H & = & (1.2 \pm 0.2) \sci{-10}. \label{eqn:wssok-abun}
\end{eqnarray}
Here $Y_P$ is the \He4 mass fraction.  These limits
restrict the present value of $\eta$ to $2.8\le\eta_{10}\le4.0$ ($\eta_{10}
\equiv 10^{10}\eta$)\@.  In what follows we will use the WSSOK values with
slight modifications to $Y_P$ and \Li7 as discussed below.

	It is noted in WSSOK that the upper limit on $Y_P$, $Y_P \le 0.24$,
may be uncertain by 0.005\@.  More recently a number of high precision
measurements of \He4 in extragalactic \hii\ regions
have been made (Pagel \etal~1992; Skillman
\etal~1994a,b; Izotov \etal~1994).  Olive and Steigman~(1994) have performed a
detailed statistical analysis of these new measurements and found the
primordial helium value
\be Y_P = 0.232 \pm 0.003 \pm 0.005, \ee
where the statistical error is listed first and the systematic error second.
The 95\% confidence range (including systematic errors) is
\be 0.221 \le Y_P \le 0.243. \label{eqn:Y-abun} \ee
We will employ this range in our analysis.

	The 95\% confidence limit quoted by WSSOK for $\Li7/\H$ consists
solely of the statistical errors in the measurements.  Recently
Thorburn~(1994) has made detailed measurements on a large number of metal poor
dwarf stars.  Her analysis employed a different model of stellar atmospheres
than the one used to derive the data compiled in WSSOK\@.  This model produces
higher effective temperatures and hence higher lithium abundances.  Her data
yield a higher mean \Li7 abundance
\be \Li7/\H = (1.8 \pm 0.1) \sci{-10} \label{eqn:li-thorburn} \ee
where the quoted error is again only the statistical uncertainty in the mean.
The difference between the \Li7 abundance given in (\ref{eqn:wssok-abun})
and (\ref{eqn:li-thorburn}) is a good estimate for the size of the
systematic errors involved in making a determination of the the
primordial \Li7 abundance. For
this work we will consider both this new upper limit and the WSSOK upper limit.

	In summary, we are using the primordial abundances of \D\ and \He3 as
found in WSSOK~(\ref{eqn:wssok-abun}) and modifications of the WSSOK values of
$Y_P$~(\ref{eqn:Y-abun}) and \Li7~(\ref{eqn:li-thorburn}) due to recent
measurements with explicit consideration of systematic errors.  The primordial
abundance limits used throughout the rest of this work are
\begin{eqnarray}
	0.221 \le & Y_P & \le 0.243, \nonumber \\
	\D/\H & \ge & 1.8\sci{-5}, \nonumber \\
	(\D + \He3)/\H & \le & 1.0\sci{-4}, \nonumber \\
	\Li7/\H & \le & 1.4\sci{-10} \quad {\rm (WSSOK)}, \nonumber \\
	 \Li7/\H & \le & 2.0\sci{-10} \quad {\rm
(Thorburn~1994)}. \label{eqn:abuns}
\end{eqnarray}

	The lithium bound is particularly important for constraining density
fluctuations.  In standard homogeneous BBN lithium must be nears its minimum
value of $\Li7/\H \sim 10^{-10}$ in order to be concordant with \D\ and $\D +
\He3$.  Obviously any density variation selects \Li7 values above the minimum,
hence \Li7 tightly constrains the range of perturbations.

\section{Model of Density Fluctuations}

	We begin with a simple model of inhomogeneities~(EP).  We assume that
some unknown process generates a baryon to photon ratio $\eta (\vec x)$ at
each point in space in such a way that the fraction of regions with a given
value $\eta$ is governed  by the distribution $f(\eta)$.  We acknowledge our
ignorance of the process that generates $\eta (\vec x)$ by assuming (instead
of deriving) the form of $f(\eta)$. Given $f(\eta)$ each region has a
constant baryon to photon ratio $\eta$ throughout nucleosynthesis.  The
regions undergo standard BBN, then mix producing the
observed abundances of light elements.  The distribution of these regions is
described in our model by the function $f(\eta)$\@.  Given the distribution
$f(\eta)$ the average mass fraction of a light element is
\be \bar X_i = \left. \int_0^\infty d\eta\, \eta f(\eta) X_i(\eta)
\right/ \bar\eta, \label{eqn:Xbar} \ee
where $X_i(\eta)$ is the mass fraction of element $i$ according to standard Big
Bang nucleosynthesis in a region with a baryon to photon ratio of $\eta$\@.
The average value of $\eta$ for the universe is given by
\be \bar\eta = \int_0^\infty d\eta\, \eta f(\eta), \ee
for $f(\eta)$ normalized,
\be \int_0^\infty d\eta\, \eta f(\eta) \equiv 1. \ee

	Shortcomings of this simple model include the assumption of equal
power in perturbations on all scales and the allowance of extremely dense
regions to contribute to the observed light element abundances today.  It has
been pointed out (Rees~1984) that high amplitude perturbations with a mass
larger than the Jean's mass at the time of recombination will form
gravitationally bound objects.  These objects prevent the baryons in them from
mixing with other baryons in the universe.  Hence these overdense regions
would not contribute to the observed abundances.

	To determine a more realistic model, we consider isocurvature
perturbations to the baryon to photon ratio.  These
perturbations are characterized by their power spectrum, \Pksq.  Given the
power spectrum the average abundance is found by
\begin{eqnarray}
  \bar X_i & = & \frac{\int \frac{d^3k}{(2\pi)^3} \Pksq \int_0^{\eta_c(k)}
d\eta\, \eta f(\eta) X_i(\eta)}{\int \frac{d^3k}{(2\pi)^3} \Pksq
\int_0^{\eta_c(k)} d\eta\, \eta f(\eta)}
\nonumber \\
           & = & \frac{\int_0^{\kmax} \frac{dk}k \Delta^2(k)
\int_0^{\eta_c(k)} d\eta\, \eta f(\eta) X_i(\eta)}{\int \frac{dk}k \Delta^2(k)
\int_0^{\eta_c(k)} d\eta\, \eta f(\eta)} \label{eqn:Xbar1}.  \end{eqnarray}
Here \be \Delta^2(k) \equiv \frac{k^3}{2\pi^2} \Pksq, \ee $\eta_c(k)$ is the
cutoff in $\eta$ based on the Jeans mass at recombination, and \kmax\ is
imposed to insure that the integrals converge.  Since the Jeans mass at
recombination, $M_J \propto \eta^{-1/2}$ (Hogan~1978; Hogan~1993) and the mass
inside a scale $k$, $M_k \propto \eta/k^3$, the cutoff in $\eta$ is \be
\eta_c(k) \propto k^2 \equiv \beta k^2, \label{eqn:eta-c} \ee where $\beta
\approx 6\sci{-14}$ for $k$ in $\Mpc^{-1}$. Using this relation and
interchanging the order of integration in $\bar X_i$~(\ref{eqn:Xbar1}) gives
\be \bar X_i = \frac{\int_0^{\etamax} d\eta\, \eta f(\eta) X_i(\eta)
\int_{(\eta/\beta)^{1/2}}^{\kmax} \frac{dk}k \Delta^2(k)}{\int_0^{\etamax}
d\eta\, \eta f(\eta) \int_{(\eta/\beta)^{1/2}}^{\kmax} \frac{dk}k
\Delta^2(k)}, \label{eqn:Xbar-scale} \ee where $\etamax \equiv \eta_c(\kmax)$.
In Eq. (\ref{eqn:Xbar-scale}) our two assumptions are manifest.  We have
imposed an upper limit on \kmax implying that there is no power in the
perturbation spectrum on scales smaller than \lambdamin$= 2\pi/$\kmax and we
have assumed that on sufficiently large scales, corresponding to $M_k > M_J$
or $k < \sqrt{\eta/\beta}$ gravitational collapse will prevent these regions
from mixing the hence these regions to not contribute in an average element
abundance. One should note however that the average value of $\eta$, ${\bar
\eta}$ is not constrained by \etamax. The density distribution indeed includes
regions with $\eta >$ \etamax, though they do not contribute to the quantities
${\bar X}_i$. We will rewrite \kmax\ in terms of \etamax\ in the rest of this
work.

\section{Results}

	For $X_i(\eta)$ we have used the standard Kawano code (Kawano~1992)
with $N_\nu=3$, $\tau_n=889\; {\rm sec}$ and the correction $\Delta Y_P =
+0.0006$ (Kernan~1993).  We begin by reviewing previous work on the gamma and
log normal distributions and provide an extended analysis of the gaussian
distribution. Then we consider a model based on inclusion of the scale of the
perturbations where the power spectrum is given by a power law.

\subsection{Gamma Distribution}

	The gamma distribution (EP, Yang~\etal~1984) is given by
\be f(\eta) = \eta^{a-1} e^{-a\eta/\bar\eta}. \ee
For this distribution the variance $\delta^2$, is
\be \delta^2 = \left( \frac{\delta\eta}{\bar\eta} \right)^2 = \frac{\<\eta^2\>
- \bar\eta^2}{\bar\eta^2} = a^{-1}. \ee
The results of varying $\delta^2$ and $\bar\eta$ are shown in
figure~\ref{fig:gamma}\@.  We have required the averaged abundances to fit the
observations~(\ref{eqn:abuns})\@.  Figure~\ref{fig:gamma} shows the abundance
contours of the light elements as given by the limits in ~(\ref{eqn:abuns})
and thus delineates the resulting
parameter space that reproduces the correct abundances. In Yang et al. (1984)
the parameters  $\delta^2$ and $\bar\eta$ were constrained to
$\bar\eta \approx 3$ and $\delta^2 \la 3$ without using the \Li7
bound and a weaker upper limit on \He4 of $Y_P < 0.25$. (For $\delta^2 < 1$,
the upper bound on $\bar\eta$ is relaxed to the homogeneous upper bound.)
Here, as one can see from the figure, the more restrictive
\He4 bound combined with the bound from D + \He3
yet with the weaker bound from \Li7 (\ref{eqn:li-thorburn})
allows us to constrain $2.8 \la \bar\eta_{10} \la 3.6$\@.
Including the WSSOK \Li7 bound this range is further constrained to $2.8 \la
\bar\eta_{10} \la 3.3$.  These results are summarized in
table~\ref{tab:eta-limits}.

\subsection{Log-Normal Distribution}

	The log-normal distribution (Barrow \& Morgan~1983) is given by \be
f(\eta) = \frac1\eta \exp \left( -\frac{(\ln\eta - \mu)^2}{2\sigma^2} \right).
\ee For this distribution \be \bar\eta = e^{\mu + \sigma^2/2}. \ee The results
of our search of parameter space are shown in figure~\ref{fig:log-normal}.
Using the weaker \Li7 bound~(\ref{eqn:li-thorburn}) then $\bar\eta_{10} \la
3.6$ is allowed. This result is similar to that given by Barrow and
Morgan~(1983), though we are using more restrictive bounds on D + \He3 and
\He4.  If we include the WSSOK \Li7 bound we are restricted to $\bar\eta_{10}
\la 3.2$\@. These results are also included in table~\ref{tab:eta-limits}.

\subsection{Gaussian Distribution}

	The gaussian distribution we consider is given by
\be f(\eta) = \exp \left( -\frac{(\eta - \mu)^2}{2\sigma^2} \right). \ee
A previous study (Sale and Mathews~1986) considered a one parameter gaussian
distribution with $\mu = 0$.
Using this two parameter distribution we find
\be \bar\eta = \mu + \sqrt{\frac2\pi} \,\sigma \frac{e^{-\mu^2/2\sigma^2}}{1 +
\erf (\mu/\sqrt 2 \sigma_.)}. \ee
Note that this expression requires
\be \bar\eta \ge \sqrt{\frac2\pi} \,\sigma. \ee
Thus a region of parameter space is already restricted by the mathematics.
The results of the parameter space search are shown
in figure~\ref{fig:gaussian}. Using the weaker \Li7
bound~(\ref{eqn:li-thorburn}) then
$\bar\eta_{10} \la 3.7$\@.  If we include the WSSOK \Li7 bound then
$\bar\eta_{10} \la 3.4$\@.  These results are summarized in
table~\ref{tab:eta-limits}.

\subsection{Power Law}

	We assume a featureless power law for the power spectrum
(Peebles~1987a,b)
\be \Pksq \propto k^n. \ee
Thus $\Delta^2(k) \propto k^{n+3}$ and the average
abundance~(\ref{eqn:Xbar-scale}) is
\be \bar X_i = \left. \int_0^{\etamax} d\eta\, \eta f(\eta) X_i(\eta)\left[ 1
- \left( \frac\eta{\etamax} \right)^{\frac{n+3}2} \right] \right/
\int_0^{\etamax} d\eta \, \eta f(\eta) \left[ 1 - \left( \frac\eta{\etamax}
\right)^{\frac{n+3}2} \right]. \label{eqn:Xbar-scale2} \ee
In the limit of $\etamax \rightarrow \infty$ ($\kmax \rightarrow \infty$) this
reduces to the simpler case where the scale of the perturbations was
ignored~(\ref{eqn:Xbar})\@.  Hence for $\etamax \gg \bar\eta$ we expect the
results to be independent of the power law index, $n$.  For each of the
previously studied distribution functions we have performed the average as
given above~(\ref{eqn:Xbar-scale2}).  The results are shown in
figs.~\ref{fig:gamma-scale1}--\ref{fig:gaussian-scale1}.  The spectral index,
$n=-0.5$ was chosen for illustrative purposes and since it is the preferred
value in the PIB model.

	As an example of the results with a power law distribution consider
the gamma distribution (fig.~\ref{fig:gamma-scale1}).  For $\delta^2=0.1$~(a)
using the weaker \Li7~(\ref{eqn:li-thorburn}) there is a small band in the
allowed parameter space between the \He4 and $\D+\He3$ bounds.  The WSSOK \Li7
value is only marginally consistent with the $\D+\He3$ bound leaving a very
narrow allowed region in parameter space.  For
$\delta^2=0.2$ (b) the limit from the weaker \Li7 value roughly overlays the
\He4 limit and the WSSOK \Li7 value is inconsistent with the $\D+\He3$ limit.
For $\delta^2=0.5$ (c) \Li7 is always produced in greater amounts than allowed
according to the WSSOK limit.  Finally for $\delta^2=1.0$~(d) \Li7 is always
overproduced and the $\D+\He3$ is inconsistent with the \He4 limit (there is
no allowed parameter space).

	From our expression for $\eta_c(k)$~(\ref{eqn:eta-c}), if $\etamax =
5\sci{-10}$ then $\lambdamin \approx 70\kpc$.  Thus for this \etamax\ the
spectrum of perturbations must be cutoff at the (comoving) scale of $70\kpc$.
No structure on scales smaller than this could be created from baryon
isocurvature baryon perturbations.  To allow smaller scale structure to form
we must lower \lambdamin\ and hence raise \etamax.  Thus we use as our
criteria for determining valid regions in parameter space that all the light
element bounds are satisfied and $\etamax \ga 5\sci{-10}$. From
fig.~\ref{fig:gamma-scale1}(a) we see that requiring structure on scales less
than $70\kpc$ eliminates the horizontal region where the light element
abundances are consistent with observations.  From
figs.~\ref{fig:gamma-scale1}(a-d), we see that this leads to a limit of
$\bar\eta_{10} \la 6.0$ with the weaker \Li7 abundance and $\bar\eta_{10} \la
4.0$ for the WSSOK \Li7 bound.  Similar results hold for the log normal
(fig.~\ref{fig:log-normal-scale1}) and gaussian
(fig.~\ref{fig:gaussian-scale1}) distributions.  These results are also
summarized in table~\ref{tab:eta-limits}.  Notice, if we allow an
arbitrary cutoff at $3.5\sci{-10} \la \etamax \la 5\sci{-10}$ we can get
$\bar\eta$ to be as large as desired.  However, this requires introducing new
physics to explain the origin of the cutoff at 70--80$\kpc$.

	We noted above that the results should be largely independent of the
power law index, $n$.  To verify this we have examined the gamma distribution
for three values of $n$.  The values chosen are $n=-0.5$
(fig.~\ref{fig:gamma-scale1}), $n = -2.5$ (fig.~\ref{fig:gamma-scale2}), and
$n = 2.5$ (fig.~\ref{fig:gamma-scale3}).  Comparing these figures we find that
decreasing $n$ slightly shifts the bounds to higher $\bar\eta$.  This effect
is most pronounced for $\delta^2=1.0$ (part~(d) of the figures).  However,
since this effect is small and shift all bound in approximately the same
manner, the limits quoted in table~\ref{tab:eta-limits} are valid for all
values of $n$.

\begin{table}[p] \center
 \begin{tabular}{p{2in}cc} \hline\hline
  Distribution & \multicolumn{2}{c}{Upper limits on $\eta_{10}$} \\ \cline{2-3}
  Function  & w/o WSSOK \Li7 & with WSSOK \Li7 \\ \hline
  Gamma & 3.6 & 3.3 \\
  Log normal & 3.6 & 3.2 \\
  Gaussian & 3.7 & 3.4 \\
  Power law + gamma & 6.0 & 4.0 \\
  Power law + log normal & 6.0 & 4.0 \\
  Power law + gaussian & 5.0 & 4.0 \\ \hline\hline
 \end{tabular}
 \caption{Upper limits on $\eta_{10}$ for various distribution functions (see
text for details).}
 \label{tab:eta-limits}
\end{table}

\section{Conclusions}

	As shown in table~\ref{tab:eta-limits} the extra parameters used in
these models of inhomogeneous BBN allow for only a slight increase in $\eta$
over SBBN\@.  This fact is easy to understand.  First ignore the \Li7 limits.
Since the abundance of \He4 is a monotonically increasing function of $\eta$
and the abundances of \D\ and \He3 are monotonically decreasing functions of
$\eta$,  when we include regions of high $\eta$ we are overproducing
(underproducing) \He4 (\D\ and \He3) in the universe.  The slight increase in
the allowed value for $\eta$ comes from the fact that \He4 is a slowly varying
function of $\eta$.  When we include the \Li7 bound it provides the tight
upper bound on $\eta$.  This is due to the fact that the observed abundance
lies in the trough of the predicted BBN production (see WSSOK)\@.  Thus any
regions of high $\eta$ greatly over produce \Li7 and we cannot allow such
regions to have a significant contribution in  the universe.  The slight
increase in $\eta$ allowed is due to the generous limits we have allowed for
the \Li7 abundance.

	The case of a power law distribution allows only a slightly extended
range at the expense of adding two new parameters, $n$ and \etamax.  The
reason for this again traces back to the above discussion of mixing in regions
with too much or too little of the light elements.  Furthermore, these results
are essentially independent of $n$.  This in turn places tight constraints on
models that contain these perturbations.  In models that allow the power
spectrum to extend to small scales the density of baryons is restricted to the
same region as homogeneous BBN\@.  Models that impose a small scale cutoff
can, if the cutoff falls in just the right region, lead to much higher values
for the baryon density at the expense of adding new physics.  Recently Gnedin,
Ostriker, \& Rees~(1994) have found some non-gaussian baryon isocurvature
models can circumvent these bounds.  We will return to this topic in a future
work.

\section*{Acknowledgments}

	We would like to thank B.~D.~Fields, G.~Gyuk, G.~Mathews,
M.~S.~Turner, P.~J.~E.~Peebles, J.~Ostriker, M.~Rees, and E. Vishniac for
useful conversations.  During the final stages of manuscript preparation we
became aware of a similar project by Jedamzik and Fuller~(1994).  Although we
have not seen their manuscript we understand they reach similar conclusions.
This work has been supported in part by NSF grant AST 90-22629, DOE grant
DE-FG02-91-ER40606, and NASA grant NAGW-1321 at the
University of Chicago, by the DOE and by NASA through grant NAGW-2381 at
Fermilab, and the U. S. Department of Energy under contract
DE-FG02-94ER-40823 at the University of Minnesota.

\endpaper

\beginapjbib

\bibitem Barrow,~J.~D. and Morgan,~J.~1983, {\em MNRAS\/} {\bf 203} 393.

\bibitem Cen,~R., Ostriker,~J.~P., and Peebles,~P.~J.~E.~1993, {\em ApJ\/}
{\bf 415} 423.

\bibitem Copi,~C.~J., Schramm,~D.~N., and Turner,~M.~S.~1994, {\em Science\/}
submitted.

\bibitem Epstein,~R.~I. and Petrosian,~V.~1975, {\em ApJ\/} {\bf 197} 281
(EP).

\bibitem Gnedin,~N.~Y., Ostriker,~J.~P., and Rees,~M.~J.~1994, {\em ApJ\/}
submitted.

\bibitem Hogan,~C.~J.~1978, {\em MNRAS\/} {\bf 185} 889.

\bibitem Hogan,~C.~J.~1993, {\em ApJ\/} {\bf 415} L63.

\bibitem Izotov,~Y.~I., Thuan,~T.~X., and Lipovetsky,~V.~A.~1994, preprint.

\bibitem Jedamzik,~K. and Fuller,~G.~M.~1994, private communication.

\bibitem Jedamzik,~K., Fuller,~G.~M., and Mathews,~G.~J.~1994, {\em ApJ\/} {\bf
423} 50.

\bibitem Kawano,~L.~H.~1992, FERMILAB-PUB-92/04-A, preprint.

\bibitem Kernan,~P.~J.~1993, The Ohio State University, PhD thesis.

\bibitem Kernan,~P,~J. \& Krauss,~L.~M.~1994, {\em PRL\/} {\bf 72} 3309.


\bibitem Kurki-Suonio,~H., Matzner,~R.~A., Olive,~K.~A., \&
        Schramm,~D.~N.~1990, {\em ApJ\/} {\bf 353} 406

\bibitem Malaney,~R.~A. and Mathews,~G.~1993, {\em Phys. Rep.} {\bf 229} 147.

\bibitem Mathews,~G.~J., Meyer,~B.~S., Alcock,~C.~R., \& Fuller,~G.~M.~1990,
        {\em ApJ\/} {\bf 358} 36.

\bibitem Olive,~K.~A. and Steigman,~G.~1994, UMN-TH-1230/94, preprint.

\bibitem Pagel,~B.~J.~E., Simonson,~E.~A., Terlevich,~R.~J., and
Edmunds,~M.~G.~1992, {\em MNRAS\/} {\bf 255} 325.

\bibitem Peebles,~P.~J.~E.~1987a, {\em Nature\/} {\bf 327} 210.

\bibitem Peebles,~P.~J.~E.~1987b, {\em ApJ\/} {\bf 315} L73.

\bibitem Rees,~M.~J.~1984, in {\em Formation and Evolution of Galaxies and
Large Structures in the Universe}, eds., J.~Audouze \& J.~Tran Thanh Van
(Boston: D.~Reidel Publishing Company), 271.

\bibitem Sale,~K.~E. and Mathews,~G.~J.~1986, {\em ApJ\/} {\bf 309} L1.

\bibitem Skillman~E.~D., Terlevich,~R.~J., Kennicut,~R.~C.,~Jr.,
Garnett,~D.~R., and Terlevich,~E~1994a, {\em ApJ\/} {\bf 431} 172.

\bibitem Skillman,~E.~D.~\etal~1994b, {\em ApJL\/} in preparation.

\bibitem Smith,~M.~S., Kawano,~L.~H., and Malaney,~R.~A.~1993, {\em ApJS\/}
{\bf 85} 219.

\bibitem Sato,~K. \& Terasawa,~N.~1991, {\em Phys. Scr.} {\bf T36}, 60

\bibitem Thomas,~D., Schramm,~D.~N., Olive,~K.~A., Mathews,~G.~J., Meyer,~B.~S.
\& Fields,~B.~D.~1994, {\em ApJ\/} {\bf 406} 569.

\bibitem Thorburn,~J.~A.~1994, {\em ApJ\/} {\bf 421} 318.

\bibitem Walker,~T.~P., Steigman,~G., Schramm,~D.~N., Olive,~K.~A., and
Kang,~H.~1991, {\em ApJ\/} {\bf 376} 51 (WSSOK).

\bibitem Yang,~J., Turner,~M.~S., Steigman,~G., Schramm,~D.~N., and
Olive,~K.~A.~1984, {\em ApJ\/} {\bf 281} 493.

\endapjbib

\beginfigures

\figure{fig:gamma} Parameter space plot for the gamma distribution.  The
acceptable parameter space is between the \He4 lines (solid), below and to the
right of the \D\ line (short-dash), above and to the right of the $\D + \He3$
line (dash-dot), and below the WSSOK \Li7 line (long-dash) or below the
Thorburn \Li7 line (long-dash, short-dash).  The light gray shaded region
satisfies the \He4, \D, $\D+\He3$, and Thorburn (weaker) \Li7 bounds.  The
dark gray shaded region satisfies the \He4, \D, $\D+\He3$, and WSSOK \Li7
bounds.

\figure{fig:log-normal} Parameter space plot for the log normal distribution.
The acceptable parameter space is as defined in figure~\ref{fig:gamma}.

\figure{fig:gaussian} Parameter space plot for the gaussian distribution.
The acceptable parameter space is as defined in figure~\ref{fig:gamma}.  The
region above the slanted solid line is not accessible for
mathematical reasons (see text for details).

\figure{fig:gamma-scale1} Parameter space plot for the gamma distribution with
scale for $n=-1/2$.  The acceptable parameter space is as defined in
figure~\ref{fig:gamma}.  In a) with $\delta^2=0.1$, $\bar\eta \la 6\sci{-10}$
for $\etamax \ga 5\sci{-10}$.  In b) with $\delta^2 = 0.2$, the WSSOK \Li7
limit does not allow any region of concordance.  In c) with $\delta^2 = 0.5$,
the Thorburn \Li7 limit falls below the $\D + \He3$ limit.  In d) with
$\delta^2 = 1.0$ \Li7 is overproduced for all values of $\bar\eta$ and
$\etamax$ and the region defined by the \D\ and $\D + \He3$ limits does not
overlap the region defined by the $Y_P$ limits.

\figure{fig:log-normal-scale1} Parameter space plot for the log normal
distribution with scale for $n=-1/2$.  The acceptable parameter space is as
defined in figure~\ref{fig:gamma}.

\figure{fig:gaussian-scale1} Parameter space plot for the gaussian
distribution with scale for $n=-1/2$.  The acceptable parameter space is as
defined in figure~\ref{fig:gamma}.

\figure{fig:gamma-scale2} Parameter space plot for the gamma
distribution with scale for $n=-2.5$.  The acceptable parameter space is as
defined in figure~\ref{fig:gamma}.

\figure{fig:gamma-scale3} Parameter space plot for the gamma
distribution with scale for $n=2.5$.  The acceptable parameter space is as
defined in figure~\ref{fig:gamma}.

\endfigures

\end{document}